# Deep Learning Generates Synthetic Cancer Histology for Explainability and Education

James M. Dolezal[1], Rachelle Wolk[2], Hanna M. Hieromnimon[1], Frederick M. Howard[1], Andrew Srisuwananukorn[3], Dmitry Karpeyev[4], Siddhi Ramesh[1], Sara Kochanny[1], Jung Woo Kwon[2], Meghana Agni[2], Richard C. Simon[2], Chandni Desai[2], Raghad Kherallah[2], Tung D. Nguyen[2], Jefree J. Schulte[5], Kimberly Cole[2], Galina Khramtsova[2], Marina Chiara Garassino[1], Aliya N. Husain[2], Huihua Li[2], Robert Grossman[6], Nicole A. Cipriani[2]*, Alexander T. Pearson[1]*

[1]Section of Hematology/Oncology, Department of Medicine, University of Chicago Medicine, Chicago, IL
[2]Department of Pathology, University of Chicago Medicine, Chicago, IL
[3]Tisch Cancer Institute, Icahn School of Medicine at Mount Sinai, New York, NY
[4]Ghost Autonomy, Inc., Mountain View, CA, USA
[5]Department of Pathology and Laboratory Medicine, University of Wisconsin at Madison, Madison, WN, USA
[6]University of Chicago, Center for Translational Data Science, Chicago, IL

* Correspondence: nicole.cipriani@bsd.uchicago.edu and alexander.pearson@bsd.uchicago.edu

## Summary

Artificial intelligence methods including deep neural networks (DNN) can provide rapid molecular classification of tumors from routine histology with accuracy that matches or exceeds human pathologists. Discerning how neural networks make their predictions remains a significant challenge, but explainability tools help provide insights into what models have learned when corresponding histologic features are poorly defined. Here, we present a method for improving explainability of DNN models using synthetic histology generated by a conditional generative adversarial network (cGAN). We show that cGANs generate high-quality synthetic histology images that can be leveraged for explaining DNN models trained to classify molecularly-subtyped tumors, exposing histologic features associated with molecular state. Fine-tuning synthetic histology through class and layer blending illustrates nuanced morphologic differences between tumor subtypes. Finally, we demonstrate the use of synthetic histology for augmenting pathologist-in-training education, showing that these intuitive visualizations can reinforce and improve understanding of histologic manifestations of tumor biology.

# Introduction

Accurate diagnosis from histopathology is the first step in the evaluation of many cancers, with management pivoting upon a tumor's morphologic classification. In addition to morphologic assessment, molecular profiling through analysis of DNA mutations, RNA fusions, and gene expression is also increasingly utilized, as a tumor's molecular subtype may inform prognosis or allow targeted therapies. Deep neural networks (DNN), a form of artificial intelligence, can classify tumors from pathologic images with high accuracy, and several studies have shown that these models can also detect actionable genetic alterations and gene expression from tumor histology even when the associated histopathological phenotype is unknown[1]. Although still in their nascent stages, DNN applications in digital pathology are being explored to assist with technical tasks, automate or augment pathologist workflows, and extend pathologist capabilities through the development of novel biomarkers[2]. DNNs are however limited by their lack of predictive transparency, which is contributing to an explainability crisis as the scientific community attempts to interpret these frequently opaque models[3,4]. When a neural network trained to detect molecular subtype performs well but the corresponding histologic features are poorly understood, explainability tools may provide insights into what the model learned and help ensure predictions are based on biologically plausible image features.

Many techniques exist for explaining artificial intelligence models in medical image analysis, as recently outlined by van der Velden, *et al*[5]. The most common explainability approaches use visual explanations that highlight areas of an image important to the final prediction. These local explainability methods, which include saliency mapping[6], attention[7], and perturbation-based approaches[8], provide insights into how a prediction was made for a specific image, contrasting with global explainability methods that provide dataset-level insights into image features learned during training. Their visual explanations are attractive due to ease of interpretability, although concerns are increasingly being raised that localizations from these techniques may not be entirely accurate for medical imaging applications[9,10]. Localization-based approaches are most helpful in instances where predictions can be attributed to a discrete object, but medical images – particularly histopathological images, which are largely textural – may not manifest predictive features amenable to clear localization. The reliability of these methods in identifying textural image features with predictive significance is unclear.

Many other approaches to providing explanations for deep learning model predictions are similarly limited for histopathological applications. Image captioning methods seek to provide text-based explanations through clearly interpretable, plain language[11], but these approaches require additional ground-truth text labels for training, and it is not clear how such approaches would translate to DNN histopathological models. Testing with concept activation vectors (TCAV) provides explanations through identifying which concepts in an image are most relevant for the prediction[12], but this approach similarly requires an additional labeled dataset and is limited to providing explanations from only prespecified concepts. Example-based explanations provide a collection of sample images similar to the specified image through analysis of neighbors in the classifier latent space[13], but do not offer insights into specific image features important to the prediction.

Generative adversarial networks (GAN) are deep learning models that use a pair of competing neural networks, called the generator and discriminator, to create realistic images. Conditional GANs (cGANs) use additional information to control the generation process, providing the ability to create images belonging to a particular class or style and smoothly transition between classes[14,15]. Recent work has shown that cGANs can be leveraged as a tool for explaining DNN classifiers, providing image-specific explanations that offer dataset-level insights into differences between image classes[16,17]. As an explainability tool, cGANs yield easily interpretable, visually clear explanations that differ from other approaches in that they are not limited to explaining localizable image features and do not require additional labeled training data.

A growing body of evidence is showing that GANs can create realistic histologic images[18–24]. Quiros *et al* showed that GANs can generate realistic artificial cancer tissue, and that traversal of the GAN latent space results in realistic images with smooth architectural changes. Levine *et al* used a cGAN to generate synthetic histologic images indistinguishable from real images, which were accurately classified by both human pathologists and DNNs trained on real images. Krause *et al* demonstrated that synthetic images generated by a cGAN improve detection of genetic alterations in CRC when used for training augmentation[21], and other groups have similarly shown that augmentation with images generated from cGANs can improve classifier performance[22,25,26]. Finally, several groups have explored the use of GANs for stain and color normalization[27–29], virtual staining[30–32], and image enhancement[33–36]. Their potential utilization as an explainability tool for histopathological DNN models, however, remains unexplored.

Here, we describe an approach to explaining histopathological DNN models using cGAN-generated synthetic histology. We show that cGAN-generated synthetic histology provides visually clear, dataset-level insights into the image features associated with DNN classifier predictions. Furthermore, we demonstrate that generation of synthetic histology can be fine-tuned through class and layer blending to provide nuanced insights into the histologic correlates for a given tumor subtype or molecular state at varying scales. Finally, we show that synthetic histologic visualizations are sufficiently intuitive and informative to improve pathology trainee classification of a rare tumor subtype.

## Results

Our approach starts with training DNN classification and cGAN models on digital pathology images (**Fig. 1A**). cGANs generate an image from the combination of a seed – a vector of random numbers that determines what the image will broadly look like – and a class label, which influences the image toward one class or another (**Fig. 1B**). The class label is converted into an embedding, a vector of numbers learned through training that encodes the essence of the class, and passed to each layer of the cGAN. For a given seed, synthetic histologic images and corresponding classifier predictions are generated for each class. If the predictions are strong and match the cGAN class label, the seed has strong classifier concordance, and if the predictions are weak but match the cGAN class label, the seed has weak concordance (**Fig. 1C**). Side-by-side, classifier-concordant image pairs illustrate histologic differences responsible for changes in classifier prediction, assisting with model explainability. To create an image in transition from one class to another, we perform a linear interpolation between two class embeddings and use the interpolated embedding for cGAN conditioning; using the same seed but gradually interpolating the embedding creates class-blended images that gradually shift from one class to another (**Fig. 1D**).

We trained a classifier and cGAN for non-small cell lung cancer (NSCLC) conditioned on adenocarcinoma vs. squamous cell carcinoma, as this is a well-described histologic phenotype suitable for assessing feasibility of the approach when the expected morphologic differences are known (**Fig. 1E** and **Supplementary Fig. 1**). Cross-validation Area Under Receiver Operator Curve (AUROC) for the classifier was 0.94 ± 0.03, with an AUROC of 0.97 on the external test set. Fréchet Inception Distance (FID)[37], a commonly used metric to evaluate realism and diversity of GAN-generated images, was 3.67 for the trained cGAN. Expert pathologist assessment revealed that strongly-concordant synthetic images were realistic and consistent with the cGAN class labels (**Supplementary Fig. 2**). The synthetic image pairs illustrated known histologic differences in adenocarcinomas and squamous cell carcinomas, including gland formation, micropapillary morphology, and papillary projections in the adenocarcinoma images, and intercellular bridging and keratinization in the squamous cell images. Some strongly concordant seeds, however, did not clearly illustrate diagnostic-grade differences between image pairs. For example, some image pairs lacked tumor, instead illustrating differences in level of necrosis, which was increased in squamous cell images, or non-diagnostic stromal changes, with an orange tint seen in some squamous cell images.

We repeated the approach for breast cancer estrogen receptor (ER) status, chosen because ER status influences morphologic phenotype on standard hematoxylin and eosin (H&E) stained slides[38,39], but the morphologic correlates are incompletely characterized (**Fig. 1F**). Cross-validation AUROC was 0.87 ± 0.02 for the classifier, with a test set AUROC of 0.81. FID for the trained cGAN was 4.46. Two expert breast pathologists concluded that the synthetic ER-negative images had higher grade, more tumor-infiltrating lymphocytes, necrosis, and/or apocrine differentiation compared with ER-positive images (**Supplementary Fig. 3**).

Finally, we trained a cGAN on thyroid neoplasms, conditioned on whether the tumor had *BRAF*$^{V600E}$-like or *RAS*-like gene expression (**Fig. 1G**). *BRAF-RAS* gene expression score, a score between -1 (*BRAF*$^{V600E}$-like) and +1 (*RAS*-like), correlates with thyroid neoplasm histologic phenotype and can be used to distinguish malignant papillary thyroid carcinomas (PTC) from the low-risk non-invasive follicular thyroid neoplasms with papillary-like nuclear features (NIFTP), despite the fact that these entities are challenging to distinguish even by experienced pathologists[40–43]. We trained a DNN regression model to predict BRS as a linear outcome and evaluated performance as a classifier by discretizing the predictions at 0, with resulting cross-validation AUROC of 0.96 ± 0.01 and external test-set AUROC of 0.98. The cGAN generated realistic and diverse images with an FID of 5.19. cGAN visualizations illustrate subtle morphologic changes associated with the *BRAF-RAS* spectrum, including nuclear changes (enlargement, chromatin clearing, membrane irregularities), architectural changes (elongated follicles, papillae), colloid changes (darkening, scalloping), and stromal changes (fibrosis, calcification, ossification) (**Figure 2**). Class blending provides realistic histologic images that gradually transition from *BRAF*$^{V600E}$-like morphology to *RAS*-like morphology (**Fig. 1H**), and predictions of these blended images smoothly change from *BRAF*$^{V600E}$-like to *RAS*-like (**Fig. 1I**).

Layer blending can provide deeper insights into class-specific morphology, as passing different embeddings to each layer in the cGAN offers a method for controlling the scale at which an image is influenced to be more like one class or another (**Fig. 3A**). For example, in **Fig. 3B**, a synthetic *RAS*-like image is shown as image B1, and a *BRAF*$^{V600E}$-like image from the same seed is shown as image B6. Passing the *BRAF*$^{V600E}$-like embedding only to layers 4-6, as shown in image B2, results in a decrease in size and variation in morphology of the follicles compared to image B1, but the prediction does not move in a *BRAF*$^{V600E}$-like direction. Passing a *BRAF*$^{V600E}$-like embedding to layers 7-9, as shown in Image B3, instead results in a minimal increase in chromatin clearing and a more eosinophilic color profile, and the classifier prediction has now moved closer to the *BRAF*$^{V600E}$-like end of the spectrum. In image B4, setting layers 10-12 to *BRAF*$^{V600E}$-like results in subtle changes to the stroma, resulting in a more ropey appearance to the collagen as well as a more eosinophilic color profile.

Finally, we tested the use of synthetic histology to augment pathologist-in-training education by creating a cGAN-based educational curriculum illustrating the *BRAF-RAS* spectrum in thyroid neoplasms (**Fig. 4**). Six pathology residents first received a standard educational lecture on thyroid neoplasms, including discussion of NIFTP subtype and differences in *BRAF*$^{V600E}$-like and *RAS*-like morphology. Residents completed a 96-question pre-test comprised of images of real tumors from a University of Chicago dataset, predicting whether images were *BRAF*$^{V600E}$-like (PTC) or *RAS*-like (NIFTP)**.** Residents then participated in a one-hour cGAN-based educational session, which included image pairs of synthetic *BRAF*$^{V600E}$-like and *RAS*-like images generated from the same seed, video interpolations showing the gradual transition from *BRAF*$^{V600E}$-like to *RAS*-like (**Supplementary Data**), and a computer-based interface in which residents could interactively generate synthetic images. Following the teaching session, residents completed a 96-question post-test comprised of real images from different cases than the pre-test. After the one-hour educational session, resident accuracy on real pathologic images significantly improved from 72.7% to 79.0% (*p* = 0.021) (**Fig. 4D**).

## Discussion

The ability to understand how deep learning histopathological classifiers make their predictions will have broad implications for the interpretability and reliability of potential DNN biomarkers and may provide an avenue for discovering how biological states manifest morphologically.

Our results show that cGANs can be leveraged as an explainability method for histopathological DNN models, providing interpretable explanations for image features associated with classifier predictions through the generation of synthetic histology. For weakly supervised histopathological applications where tile-level image labels are inherited from the parent slide, not all labeled images in the training dataset are expected to possess image features associated with the outcome of interest, such as image tiles with background normal tissue, pen marks, out-of-focus areas, or artifact. GANs are trained to generate images from the entire training distribution, which includes these potentially uninformative image tiles. By filtering cGAN-generated images with classifier concordance, we use the classifier's predictions to identify synthetic image pairs that are enriched for morphologic differences related to the outcome of interest. We demonstrate that this approach highlights known morphologic differences between lung adenocarcinoma and squamous cell carcinomas, ER-negative and ER-positive breast cancers, and *BRAF*$^{V600E}$-like and *RAS*-like thyroid neoplasms. Generating class-blended images which smoothly transition from one class to another through linear interpolation of the embedding latent space further improves intuitiveness of the synthetic image explanations.

This approach has some key advantages over other explainability methods for histopathological models. It can illustrate subtle morphologic differences which do not easily localize within an image, such as variations in hue and color, staining differences, and changes in stromal characteristics. It also does not require *a priori* specification of captions or concepts thought to be important for prediction, capturing a broader array of morphologic differences and permitting more exploratory use. Finally, it does not require any additional labeled training data, instead utilizing the same training distribution used for the classifier being explained. When combined with other explainability approaches, class- and layer-blending with cGANs can improve richness of classifier explanations and support biological plausibility of DNN predictions.

cGAN-generated synthetic histology also provides an avenue for hypothesis generation through illustration of morphologic patterns capable of being learned by deep learning. The application of this method in thyroid neoplasms is a particularly salient example: a DNN trained to predict a gene expression signature and a cGAN conditioned on the same signature together illustrate the morphologic manifestation of *BRAF*$^{V600E}$-like and *RAS*-like gene expression. This not only provides a method of explaining the DNN classifier – it also provides insights into how underlying tumor biology is connected with morphologic phenotype. A similar approach could be used to investigate morphologic manifestations of other molecular states in any cancer.

In addition to its use for DNN explainability, this approach provides a potential tool for pathology trainee education, particularly for rare tumors or elusive diagnoses. cGANs can generate synthetic histology that, with curation by an expert pathologist, depict subtle differences significant for diagnosis that may otherwise be challenging to clearly demonstrate with real histopathological images. The generation process can be both controlled and fine-tuned, allowing an educator to build a curriculum using synthetic histologic images for a particular objective or phenomenon as a supplement to real images. In our small study with pathology residents at a single institution, a short education curriculum utilizing synthetic histology improved trainee recognition of diagnostically challenging thyroid neoplasms.

Although synthetic histology offers compelling possibilities for both DNN explainability and trainee education, some important limitations must be acknowledged. As an explainability tool, cGANs provide global explanations using synthetic image examples. Thus, their utility is in improving understanding at the dataset level, rather than providing local insights into why a prediction was made for a specific real image. It also requires training a separate GAN model, incurring additional computational time. Finally, GANs will perpetuate any underlying biases in the training dataset. This is advantageous in the context of explainability, as it may assist with highlighting potential biasing factors such as stain or color differences. Careful curation by an expert pathologist will be required to utilize synthetic histology for education in order to prevent perpetuation of potential biases in the training dataset.

In summary, cGANs can generate realistic, class-specific histologic images, and exploring visualizations from images with high classifier concordance provides an intuitive tool for deep learning model explainability. Class blending via embedding interpolation yields realistic images with smooth transitions between classes, and layer blending reveals unique morphological constructs at architectural, cellular, and stromal levels. Synthetic histology not only offers an approach to model explainability, but can also provide new, hypothesis-generating insights into histologic associations with molecularly-defined tumor subtypes. Finally, synthetic histology can also be an effective teaching aid, capable of improving trainee recognition of histologic classes in a rare cancer subtype.

## Methods

### Dataset description

The Lung cGAN was trained on 941 whole-slide images (WSI) from The Cancer Genome Atlas (TCGA), including 467 slides from the lung adenocarcinoma project (TCGA-LUAD) and 474 slides from the lung squamous cell carcinoma project (TCGA-LUSC) (https://portal.gdc.cancer.gov/). Validation was performed on 1,306 WSIs from the Clinical Proteomic Tumor Analysis Consortium (CPTAC) lung adenocarcinoma (CPTAC-LUAD) and lung squamous cell carcinoma (CPTAC-LSCC) collections (https://www.cancerimagingarchive.net/collections/). The Breast cGAN was trained on 1,048 WSIs from The Cancer Genome Atlas (TCGA-BRCA), including 228 estrogen receptor (ER) negative tumors and 820 ER-positive tumors. Validation was performed on 98 WSIs from CPTAC, including 26 ER-negative and 72 ER-positive tumors. The Thyroid cGAN was trained on 369 WSIs from The Cancer Genome Atlas (TCGA-THCA), including 116 $BRAF^{V600E}$-like tumors (where $BRAF$-$RAS$ gene expression score is less than 0) and 271 $RAS$-like tumors (where $BRAF$-$RAS$ gene expression score is greater than 0). Validation was performed on an institutional dataset of 134 tumors, including 76 $BRAF^{V600E}$-like PTCs and 58 $RAS$-like NIFTPs.

### Image processing

For the classifier models, image tiles were extracted from WSIs with an image tile width of 302 μm and 299 pixels using Slideflow version 1.3.1[44]. For the breast and lung cGANs, image tiles were extracted with an image tile width of 400 μm and 512 pixels. For the thyroid cGAN, image tiles were extracted at 302 μm and 512 pixels. Background was removed via grayspace filtering, Otsu's thresholding, and gaussian blur filtering. Gaussian blur filtering was performed with a sigma of 3 and threshold of 0.02. Image tiles were extracted from within pathologist-annotated regions of interest (ROIs) to maximize cancer-specific training.

### Classifier training

We trained deep learning classification models based on an Xception architecture, using ImageNet pretrained weights and two hidden layers of width 1024, with dropout ($p$ = 0.1) after each hidden layer. Models were trained with Slideflow using the Tensorflow backend with a single set of hyperparameters and category-level mini-batch balancing (**Supplementary Table 1**). Training images were augmented with random flipping and cardinal rotation, JPEG compression (50% chance of compression with quality level between 50-100%), and gaussian blur (10% chance of blur with sigma between 0.5-2.0). Binary categorization models (lung and breast classifiers) were trained with cross-entropy loss, and the thyroid BRS classifier was trained with mean squared error loss. Models are first trained with site-preserved cross-validation[45], then a final model is trained across the full dataset and validated on an external dataset. Classifier models were evaluated by calculating Area Under Receiver Operator Curve (AUROC), with cross-validation AUROC reported as mean ± SD.

### cGAN training

Our cGAN architecture is an implementation of StyleGAN2, minimally modified to interface with the histopathology deep learning package Slideflow and allow for easier embedding space interpolation[14]. The lung cGAN was conditioned on the binary category of adenocarcinoma vs. squamous cell carcinoma, and the breast cGAN was conditioned on the binary category of ER-negative vs. ER-positive. The thyroid cGAN was conditioned on a binary categorization of the continuous *BRAF-RAS* score, discretized at 0 into *BRAF*[V600E]-like (less than 0) or *RAS*-like (greater than 0).

The lung cGAN was trained on 4 A100 GPUs for 25,000 kimg (25 million total images) starting with random weight initialization. The breast cGAN was trained on 4 A100 GPUs for 10,000 kimg (10 million total images), and the thyroid cGAN was trained on 2 A100 GPUs for 12,720 kimg (12.7 million total images), stopped at this time point due to model divergence with further training. All cGANs were trained with an R1 gamma of 1.6384, batch size of 32, and using all available augmentations from StyleGAN2. cGANs were evaluated by calculating Fréchet Inception Distance (FID)[37] using the full real image dataset and 50,000 cGAN-generated images.

### *Classifier concordance*

To assess classifier concordance for a cGAN and an associated classifier, the cGAN generates an image for each class using the same seed. The generated images are center-cropped and resized to match the same histologic magnification as the associated classifier, and the classifier creates predictions for each image. Predictions are considered "strong" if the post-softmax value is greater than 0.75 for the predicted class, and "weak" if the post-softmax value for the predicted class is less than 0.75. For the thyroid BRS classifier which uses a continuous outcome, predictions are considered "strong" if the raw prediction is less than -0.5 or greater than 0.5, and "weak" if the prediction is between -0.5 and 0.5. A given seed is defined as strongly concordant if the classifier predictions match the cGAN class labels for both images and the predictions are both strong. A seed is weakly concordant if the classifier predictions match the cGAN class labels, but either prediction is weak. A seed is non-concordant if the classifier predictions do not match the cGAN class labels.

### *cGAN class and layer blending*

To create class-conditional images, cGAN class labels are projected into an embedding space before conditioning the network, with the projection learned during training. After training, each class label has a single associated embedding vector. To create class-blended images, we perform a linear interpolation between class embeddings and use these interpolated embeddings for network conditioning while holding the cGAN seed constant. We create layer-blended images by passing different class embeddings to each cGAN network layer while holding the cGAN seed constant.

### *Pathologist assessment of cGAN images*

Domain-expert pathologists reviewed at least 50 strongly-concordant synthetic histologic images to assess realism, variety, and consistency with cGAN class labels. Pathologists first reviewed the images in a blinded fashion without knowledge of the associated cGAN labels. Lossless, PNG images were viewed at the full 512 x 512 px resolution. Pathologists then reviewed the strongly-concordance synthetic image pairs side-by-side with knowledge of the cGAN labels to assess consistency of the synthetic images with biological expectations for the associated class labels. Pathologists described histologic differences between each image pairs and provided an overall summary of thematic differences between classes.

### *cGAN educational session*

Six pathology were recruited for this study via email. No sample-size calculation was performed prior to recruitment. Participating residents received a one-hour lecture as a part of their core educational curriculum discussing the histopathological diagnosis of thyroid neoplasms, including a discussion of differentiating between malignant papillary thyroid carcinomas (PTCs), including follicular-variant PTCs, and benign non-invasive follicular thyroid neoplasms with papillary-like nuclear features (NIFTP). A discussion of the molecular association between PTCs and *BRAF*[V600E] mutations, and NIFTPs and *RAS* mutations, was also included.

Pathology residents then took a pre-test based on 96 real images from 48 cases at the University of Chicago, including 24 PTCs (both classic and follicular-variant) and 24 NIFTPs. The trained BRS classifier model generated predictions across all whole-slide images, and for each case, three strongly-predicted image tiles (prediction less than -0.5 or greater than 0.5) were randomly selected and merged side-by-side, and three weakly-predicted image tiles (prediction between -0.5 and 0.5) were randomly selected and merged, resulting in two merged image trios for each of the 48 cases. The pre-test was comprised of weak and strong image trios for 24 PTCs and NIFTPs, and residents were asked to predict whether the image trios came from a $BRAF^{V600E}$-like tumor (PTC) or $RAS$-like tumor (NIFTP).

Residents then participated in a one-hour cGAN-based educational curriculum. The curriculum was developed by first calculating classifier concordance for 1000 seeds and identifying the strongly-concordant seeds. $BRAF^{V600E}$-like and $RAS$-like image pairs for the 412 strongly-concordant seeds were reviewed by a domain expert pathologist, and 46 were chosen for inclusion in the teaching session. Video interpolations were generated and shown for seven of these seeds. The educational session was structured as a PowerPoint presentation, using only synthetic histologic image pairs and video interpolations to highlight important morphologic differences associated with the BRS spectrum. Residents also had access to a computer workstation loaded with an interactive visualization of cGAN generated images and class blending, to supplement the visualizations shown in slideshow format.

Finally, residents completed a post-test comprised of 96 images from 48 different cases than the pre-test, and resident classification accuracy was compared using a one-sided paired T-test.

## Data availability

All data and associated accession numbers used for training is included in this repository, and can be additionally accessed directly at https://portal.gdc.cancer.gov/ and https://www.cancerimagingarchive.net/collections/ using the accession numbers provided in the **Supplementary Data**. Restrictions apply to the availability of the internal University of Chicago thyroid dataset, but all requests will be promptly evaluated based on institutional and departmental policies to determine whether the data requested are subject to intellectual property or patient privacy obligations. The University of Chicago dataset can only be shared for non-commercial academic purposes and will require a data user agreement.

## Code availability

All code and models are made publicly available with an interactive user interface for class blending and latent space navigation at https://github.com/jamesdolezal/synthetic-histology. The user interface provided is the same interface used during the educational teaching session.

## Ethics statement

Educational study was reviewed by institutional IRB and deemed minimal risk, exempt from protocol approval and requirement for informed consent.

## Acknowledgements

This work was funded by the National Institute of Health / National Cancer Institute NIH/NCI) U01-CA243075 (ATP), National Institute of Health / National Institute of Dental and Craniofacial Research (NIH/NIDCR) R56-DE030958 (ATP), grants from Cancer Research Foundation (ATP), grants from Stand Up to Cancer (SU2C) Fanconi Anemia Research Fund – Farrah Fawcett Foundation Head and Neck Cancer Research Team Grant (ATP), Horizon 2021-SC1-BHC I3LUNG grant (ATP, MCC), and grants from ECOG Research and Education Foundation (AS).

## Declaration of interests

DK is an employee of Ghost Autonomy, Inc., but holds no competing interests for this work. ATP reports personal fees from Prelude Therapeutics Advisory Board, personal fees from Elevar Advisory Board, personal fees from AbbVie consulting, and personal fees from Ayala Advisory Board, all outside of submitted work. ATP reports stock options ownership in Privo Therapeutics. All remaining authors report no competing interests.

## Author contributions

J.M.D., N.A.C., and A.T.P. conceived the study. J.M.D. wrote the code and performed the experiments. J.M.D., H.M.H., F.M.H., A.S., D.K., S.R., and S.K analyzed data. R.W. and N.A.C. created and taught the educational curriculum. J.W.K., M.A., R.C.S., C.D., R.K., and T.D.N. assessed synthetic histology image quality and participated in the educational curriculum. J.J.S provided annotated regions of interest. R.W., K.C., G.K., A.N.H., H.L., and N.A.C. performed expert pathologist review of synthetic histology images. J.M.D wrote the manuscript. M.C.G. and R.G. provided valuable comments on the general approach. All authors contributed to discussion and revision of the manuscript.

## Figure Legends

*Figure 1.* **cGANs illustrate molecular states expressed in tumor histopathology with synthetic histology.** **(a)** For a given set of training data, two models are trained – a conditional generative adversarial network (cGAN), which generates images, and a classification model, which predicts histologic class from an image. **(b)** cGANs create synthetic images from a seed of random noise and a class label, passed to each layer in the network through an embedding. Synthetic images are then classified by the trained classifier. **(c)** For a given seed, synthetic images are generated for each histologic class, and the classifier creates predictions for each image. Seeds for which predictions match the GAN labels are classifier-concordant. Visualizing classifier-concordant image pairs side-by-side allows one to appreciate the histologic features responsible for classifier predictions, providing a tool for model explainability and education. **(d)** Class-blended images are generated by interpolating between class embeddings while holding the seed constant. **(e)** A cGAN was trained on lung adenocarcinoma vs. squamous cell carcinoma. Classifier concordance for 1000 seeds was 31.1% strong, 27.0% weak, and 41.9% non-concordant. (**f**) A second cGAN was trained on breast cancer estrogen receptor (ER) status determined by immunohistochemistry (IHC). Classifier concordance for 1000 seeds was 25.9% strong, 10.0% weak, and 64.1% non-concordant. (**g**) A final cGAN was trained on thyroid neoplasm *BRAF-RAS* gene expression score (BRS). Classifier concordance for 1000 seeds was 41.2% strong, 36.2% weak, and 22.6% non-concordant. **(h)** Illustration of class blending performed for a strongly-concordant Thyroid cGAN seed, transitioning from *BRAF*$^{V600E}$-like to *RAS*-like. **(i)** Example class-blended images for the Lung and Breast cGANs. **(j)** Predictions smoothly transition during class blending.

*Figure 2.* **cGAN-generated synthetic histology illustrates morphologic differences associated with *BRAF-RAS* gene expression score in thyroid neoplasms.** Classifier-concordant seeds from the thyroid cGAN were reviewed with an expert thyroid pathologist and pathology fellow to determine thematic differences in cGAN-generated *BRAF*$^{V600E}$-like and *RAS*-like histologic features. Seed 0 illustrates architectural differences, with a papillae in the *BRAF*$^{V600E}$-like image replaced with colloid in the *RAS*-like image. Seed 3 highlights the an increase in fibrosis in the *BRAF*$^{V600E}$-like image. The *BRAF*$^{V600E}$-like image for seed 12 shows a cystic structure with cell lining, which is replaced with what appears to be a tear in the *RAS*-like image, accompanied by architectural differences moving from papillae in the *BRAF*$^{V600E}$-like image to follicles in the *RAS*-like image. Seed 16 demonstrates an increase in lacunae caused by resorbed colloid in the *BRAF*$^{V600E}$-

like image, compared with smaller, more regular follicles in the *RAS*-like image. Seed 18 shows papillae, a papillary vessel, and a cystic structure in the *BRAF*$^{V600E}$-like image replaced with follicles, colloid, and an endothelial-lined vessel in the *RAS*-like image, respectively. Seed 30 highlights more tumor-infiltrating lymphocytes in the *BRAF*$^{V600E}$-like along with increased cytoplasmic density compared with the *RAS*-like image. Seed 35 shows overall similar architecture in the two images, but with greater cell flattening in the *RAS*-like image compared to the *BRAF*$^{V600E}$-like image. Seeds 102 and 128 both illustrate nuclear pleomorphism, increased cytoplasm, papillae, and scalloping in the *BRAF*$^{V600E}$-like images compared with the *RAS*-like image. Seed 167 highlights nuclear pleomorphism and fibrosis in the *BRAF*$^{V600E}$-like image compared with *RAS*-like image which has monotonous, circular, non-overlapping nuclei with regular contours and fine, dark chromatin.

*Figure 3*. **Class and layer blending provides nuanced insights into *BRAF*$^{V600E}$-like and *RAS*-like thyroid neoplasms. (a)** cGANs can create synthetic layer-blended images by conditioning the network using different embeddings at each layer. **(b)** Layer blending with a seed from the Thyroid cGAN reveals different morphologic changes associated with the *RAS*-like and *BRAF*$^{V600E}$-like gene expression spectrum. Each image includes a corresponding classifier prediction, from -1 (*BRAF*$^{V600E}$-like) to +1 (*RAS*-like). Image B1 is a fully *RAS*-like image, and Image B6 is a fully *BRAF*$^{V600E}$-like image. Images B2-B5 are generated by using different class embeddings at each cGAN layer. Examining the resulting morphologic changes that occur when passing the *BRAF*$^{V600E}$-like embedding to different layers illustrates different types of morphologic changes associated with the *BRAF*$^{V600E}$-RAS spectrum.

*Figure 4*. **Synthetic histology augments pathologist-in-training education. (a)** Schematic for creating GAN-based educational curriculum. A trained classifier and cGAN were used to generate and select classifier-concordant synthetic histology images, along with class blending videos generated through embedding interpolation. Synthetic histology images and videos were curated by an expert pathologist and incorporated into an educational curriculum. **(b)** Schematic for assessing effect of cGAN-based educational session on ability for pathology trainees to accurate classify images from real thyroid neoplasms. 48 *BRAF*$^{V600E}$-like PTCs and 48 RAS-like NIFTPs were randomly split into a pre-test and post-test dataset. Predictions were generated for tiles from each slide using the trained BRS classifier and separated into weakly correct predictions (correct predictions between -0.5 and 0.5) and strongly correct predictions (correct predictions <-0.5 or >0.5). For each slide, three weakly-correct images were randomly selected and merged into a single image trio, and three strongly-correct images were randomly selected and merged, resulting in 2 images per slide. The pre-test thus consisted of 96 images from 48 slides. The same procedure was taken for the post-test, comprised of 96 images from 48 different slides. **(c)** Example real image trios used during pre-test or post-test. **(d)** Trainee classification accuracy on real images significantly improved after the teaching session, from 72.7% to 79.0% (*p* = 0.021). **(e)** Improvement in trainee classification accuracy was greater for real images with strong classifier predictions (74.3% to 83.0%, *p* = 0.012) compared to real images with weak predictions (70.8% to 75.0%, *p* = 0.132).

## References

1. Fu, Y. *et al.* Pan-cancer computational histopathology reveals mutations, tumor composition and prognosis. *Nat. Cancer* **1**, 800–810 (2020).
2. Heinz, C. N., Echle, A., Foersch, S., Bychkov, A. & Kather, J. N. The future of artificial intelligence in digital pathology – results of a survey across stakeholder groups. *Histopathology* **80**, 1121–1127 (2022).
3. Ghassemi, M., Oakden-Rayner, L. & Beam, A. L. The false hope of current approaches to explainable artificial intelligence in health care. *Lancet Digit. Health* **3**, e745–e750 (2021).
4. Reddy, S. Explainability and artificial intelligence in medicine. *Lancet Digit. Health* **4**, e214–e215 (2022).
5. van der Velden, B. H. M., Kuijf, H. J., Gilhuijs, K. G. A. & Viergever, M. A. Explainable artificial intelligence (XAI) in deep learning-based medical image analysis. *Med. Image Anal.* **79**, 102470 (2022).

6. Zeiler, M. D. & Fergus, R. Visualizing and Understanding Convolutional Networks. in *Computer Vision – ECCV 2014* (eds. Fleet, D., Pajdla, T., Schiele, B. & Tuytelaars, T.) 818–833 (Springer International Publishing, 2014).
7. Jetley, S., Lord, N. A., Lee, N. & Torr, P. H. S. Learn To Pay Attention. (2018) doi:10.48550/ARXIV.1804.02391.
8. Fong, R. C. & Vedaldi, A. Interpretable Explanations of Black Boxes by Meaningful Perturbation. in *Proceedings of the IEEE International Conference on Computer Vision (ICCV)* (2017).
9. Arun, N. *et al.* Assessing the Trustworthiness of Saliency Maps for Localizing Abnormalities in Medical Imaging. *Radiol. Artif. Intell.* **3**, e200267 (2021).
10. Saporta, A. *et al.* Benchmarking saliency methods for chest X-ray interpretation. *Nat. Mach. Intell.* **4**, 867–878 (2022).
11. Vinyals, O., Toshev, A., Bengio, S. & Erhan, D. Show and Tell: A Neural Image Caption Generator. in *Proceedings of the IEEE Conference on Computer Vision and Pattern Recognition (CVPR)* (2015).
12. Kim, B. *et al.* Interpretability beyond feature attribution: Quantitative Testing with Concept Activation Vectors (TCAV). *35th International Conference on Machine Learning, ICML 2018* vol. 6 4186–4195 (2018).
13. Uehara, K., Murakawa, M., Nosato, H. & Sakanashi, H. Prototype-Based Interpretation of Pathological Image Analysis by Convolutional Neural Networks. in *Pattern Recognition* (eds. Palaiahnakote, S., Sanniti di Baja, G., Wang, L. & Yan, W. Q.) 640–652 (Springer International Publishing, 2020).
14. Karras, T. *et al.* Analyzing and Improving the Image Quality of StyleGAN. in *Proc. CVPR* (2020).
15. T. Karras, S. Laine, & T. Aila. A Style-Based Generator Architecture for Generative Adversarial Networks. in *2019 IEEE/CVF Conference on Computer Vision and Pattern Recognition (CVPR)* 4396–4405 (2019). doi:10.1109/CVPR.2019.00453.
16. Lang, O. *et al.* Explaining in Style: Training a GAN to explain a classifier in StyleSpace. (2021) doi:10.48550/ARXIV.2104.13369.
17. Shih, S.-M., Tien, P.-J. & Karnin, Z. S. GANMEX: One-vs-One Attributions using GAN-based Model Explainability. in *ICML* (2021).
18. Levine, A. B. *et al.* Synthesis of diagnostic quality cancer pathology images by generative adversarial networks. *J. Pathol.* **252**, 178–188 (2020).
19. McAlpine, E., Michelow, P., Liebenberg, E. & Celik, T. Is it real or not? Toward artificial intelligence-based realistic synthetic cytology image generation to augment teaching and quality assurance in pathology. *J. Am. Soc. Cytopathol.* **11**, 123–132 (2022).
20. Zhao, J., Hou, X., Pan, M. & Zhang, H. Attention-based generative adversarial network in medical imaging: A narrative review. *Comput. Biol. Med.* **149**, 105948 (2022).
21. Krause, J. *et al.* Deep learning detects genetic alterations in cancer histology generated by adversarial networks. *J. Pathol.* **254**, 70–79 (2021).
22. Deshpande, S., Minhas, F., Graham, S. & Rajpoot, N. SAFRON: Stitching Across the Frontier Network for Generating Colorectal Cancer Histology Images. *Med. Image Anal.* **77**, 102337 (2022).
23. Tschuchnig, M. E., Oostingh, G. J. & Gadermayr, M. Generative Adversarial Networks in Digital Pathology: A Survey on Trends and Future Potential. *Patterns* **1**, 100089 (2020).
24. Quiros, A. C., Murray-Smith, R. & Yuan, K. PathologyGAN: Learning deep representations of cancer tissue. in *Proceedings of the Third Conference on Medical Imaging with Deep Learning* (eds. Arbel, T. et al.) vol. 121 669–695 (PMLR, 2020).
25. Chen, R. J., Lu, M. Y., Chen, T. Y., Williamson, D. F. K. & Mahmood, F. Synthetic data in machine learning for medicine and healthcare. *Nat. Biomed. Eng.* **5**, 493–497 (2021).
26. Wei, J. *et al.* Generative Image Translation for Data Augmentation in Colorectal Histopathology Images. in *Proceedings of the Machine Learning for Health NeurIPS Workshop* (eds. Dalca, A. V. et al.) vol. 116 10–24 (PMLR, 2020).
27. F. G. Zanjani, S. Zinger, B. E. Bejnordi, J. A. W. M. van der Laak, & P. H. N. de With. Stain normalization of histopathology images using generative adversarial networks. in *2018 IEEE 15th International Symposium on Biomedical Imaging (ISBI 2018)* 573–577 (2018). doi:10.1109/ISBI.2018.8363641.
28. M. T. Shaban, C. Baur, N. Navab, & S. Albarqouni. Staingan: Stain Style Transfer for Digital Histological Images. in *2019 IEEE 16th International Symposium on Biomedical Imaging (ISBI 2019)* 953–956 (2019). doi:10.1109/ISBI.2019.8759152.

29. Lafarge, M. W., Pluim, J. P. W., Eppenhof, K. A. J., Moeskops, P. & Veta, M. Domain-Adversarial Neural Networks to Address the Appearance Variability of Histopathology Images. in *Deep Learning in Medical Image Analysis and Multimodal Learning for Clinical Decision Support* (eds. Cardoso, M. J. et al.) 83–91 (Springer International Publishing, 2017).
30. Erik A. Burlingame, Adam A. Margolin, Joe W. Gray, & Young Hwan Chang. SHIFT: speedy histopathological-to-immunofluorescent translation of whole slide images using conditional generative adversarial networks. in vol. 10581 1058105 (2018).
31. Xu, Z., Moro, C. F., Bozóky, B. & Zhang, Q. GAN-based Virtual Re-Staining: A Promising Solution for Whole Slide Image Analysis. *ArXiv* **abs/1901.04059**, (2019).
32. Bayramoglu, N., Kaakinen, M., Eklund, L. & Heikkila, J. Towards Virtual H&E Staining of Hyperspectral Lung Histology Images Using Conditional Generative Adversarial Networks. in *Proceedings of the IEEE International Conference on Computer Vision (ICCV) Workshops* (2017).
33. B. Venkatesh, T. Shaht, A. Chen, & S. Ghafurian. Restoration of Marker Occluded Hematoxylin and Eosin Stained Whole Slide Histology Images Using Generative Adversarial Networks. in *2020 IEEE 17th International Symposium on Biomedical Imaging (ISBI)* 591–595 (2020). doi:10.1109/ISBI45749.2020.9098358.
34. Çelik, G. & Talu, M. F. Resizing and cleaning of histopathological images using generative adversarial networks. *Phys. Stat. Mech. Its Appl.* **554**, 122652 (2020).
35. Upadhyay, U. & Awate, S. P. A Mixed-Supervision Multilevel GAN Framework for Image Quality Enhancement. in *Medical Image Computing and Computer Assisted Intervention – MICCAI 2019* (eds. Shen, D. et al.) 556–564 (Springer International Publishing, 2019).
36. F. Shahidi. Breast Cancer Histopathology Image Super-Resolution Using Wide-Attention GAN With Improved Wasserstein Gradient Penalty and Perceptual Loss. *IEEE Access* **9**, 32795–32809 (2021).
37. Heusel, M., Ramsauer, H., Unterthiner, T., Nessler, B. & Hochreiter, S. GANs Trained by a Two Time-Scale Update Rule Converge to a Local Nash Equilibrium. in *Proceedings of the 31st International Conference on Neural Information Processing Systems* 6629–6640 (Curran Associates Inc., 2017).
38. Naik, N. *et al.* Deep learning-enabled breast cancer hormonal receptor status determination from base-level H&E stains. *Nat. Commun.* **11**, 5727 (2020).
39. Couture, H. D. *et al.* Image analysis with deep learning to predict breast cancer grade, ER status, histologic subtype, and intrinsic subtype. *NPJ Breast Cancer* **4**, 30 (2018).
40. Dolezal, J. M. *et al.* Deep learning prediction of BRAF-RAS gene expression signature identifies noninvasive follicular thyroid neoplasms with papillary-like nuclear features. *Mod. Pathol.* **34**, 862–874 (2021).
41. Elsheikh, T. M. *et al.* Interobserver and Intraobserver Variation Among Experts in the Diagnosis of Thyroid Follicular Lesions With Borderline Nuclear Features of Papillary Carcinoma. *Am. J. Clin. Pathol.* **130**, 736–744 (2008).
42. Hirokawa, M. *et al.* Observer variation of encapsulated follicular lesions of the thyroid gland. *Am. J. Surg. Pathol.* **26**, 1508–1514 (2002).
43. Lloyd, R. V. *et al.* Observer variation in the diagnosis of follicular variant of papillary thyroid carcinoma. *Am. J. Surg. Pathol.* **28**, 1336–1340 (2004).
44. Dolezal, J., Kochanny, S. & Howard, F. Slideflow: A Unified Deep Learning Pipeline for Digital Histology. (2022) doi:10.5281/zenodo.7301864.
45. Howard, F. M. *et al.* The impact of site-specific digital histology signatures on deep learning model accuracy and bias. *Nat. Commun.* **12**, 4423 (2021).

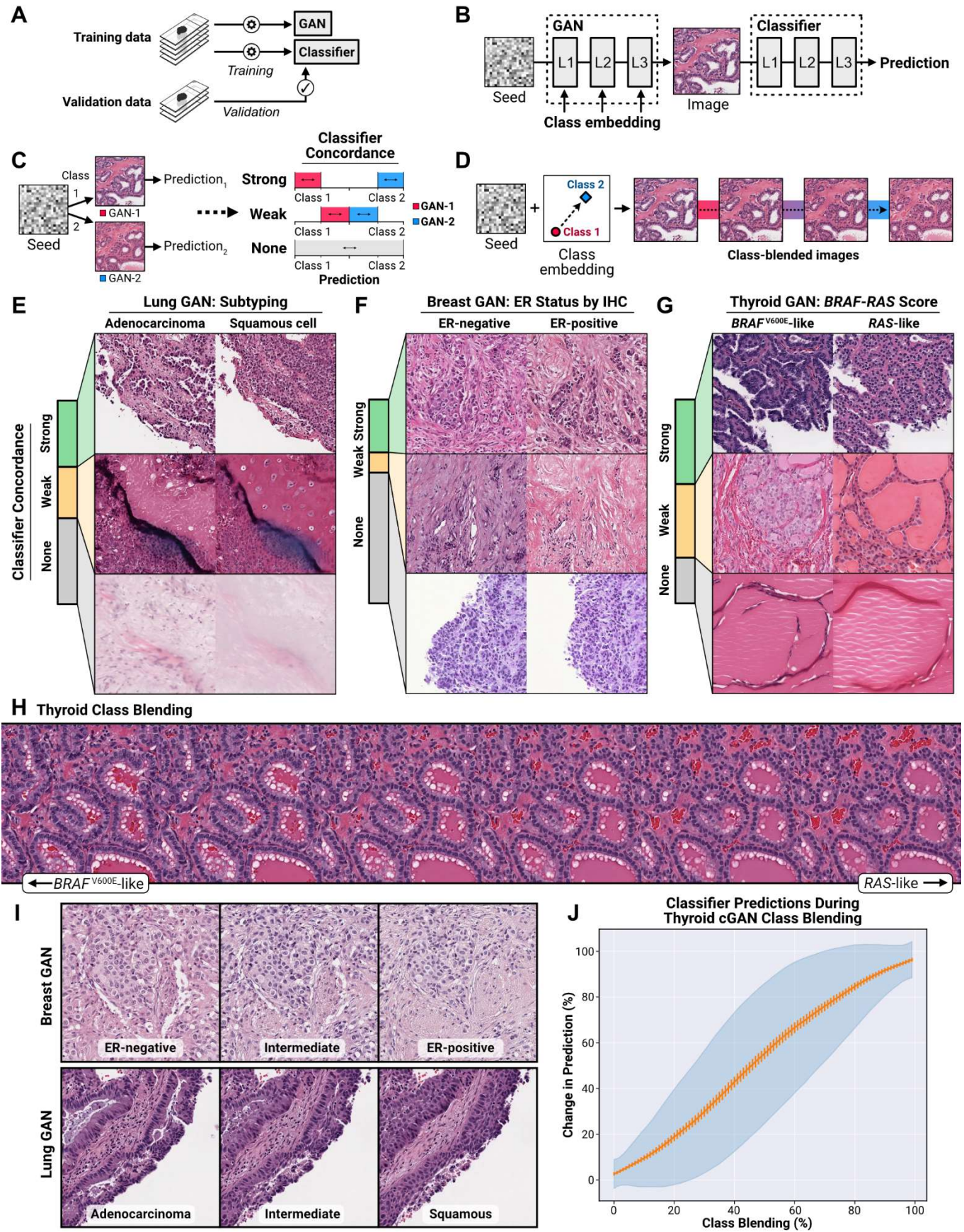

**Figure 1**

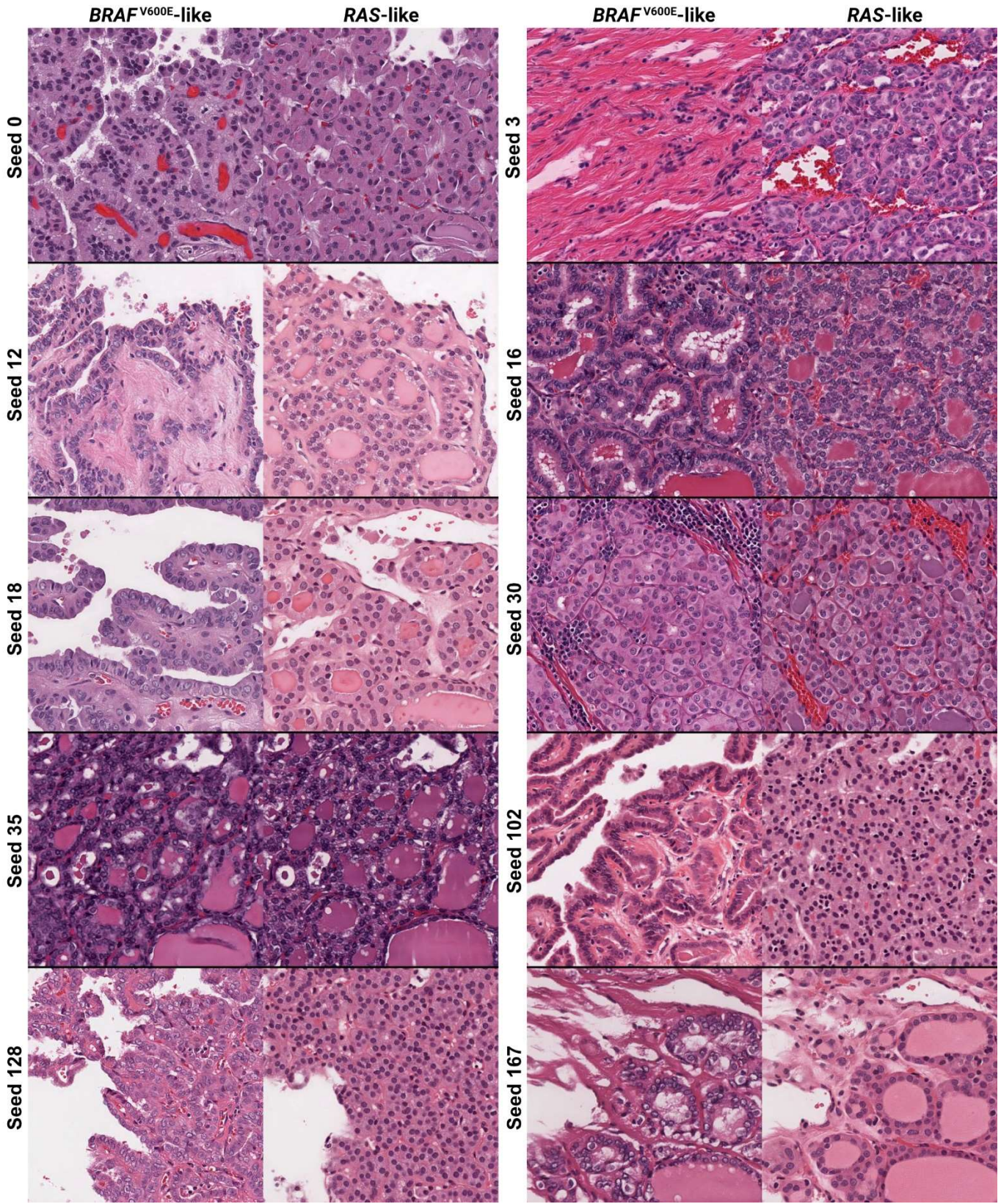

**Figure 2**

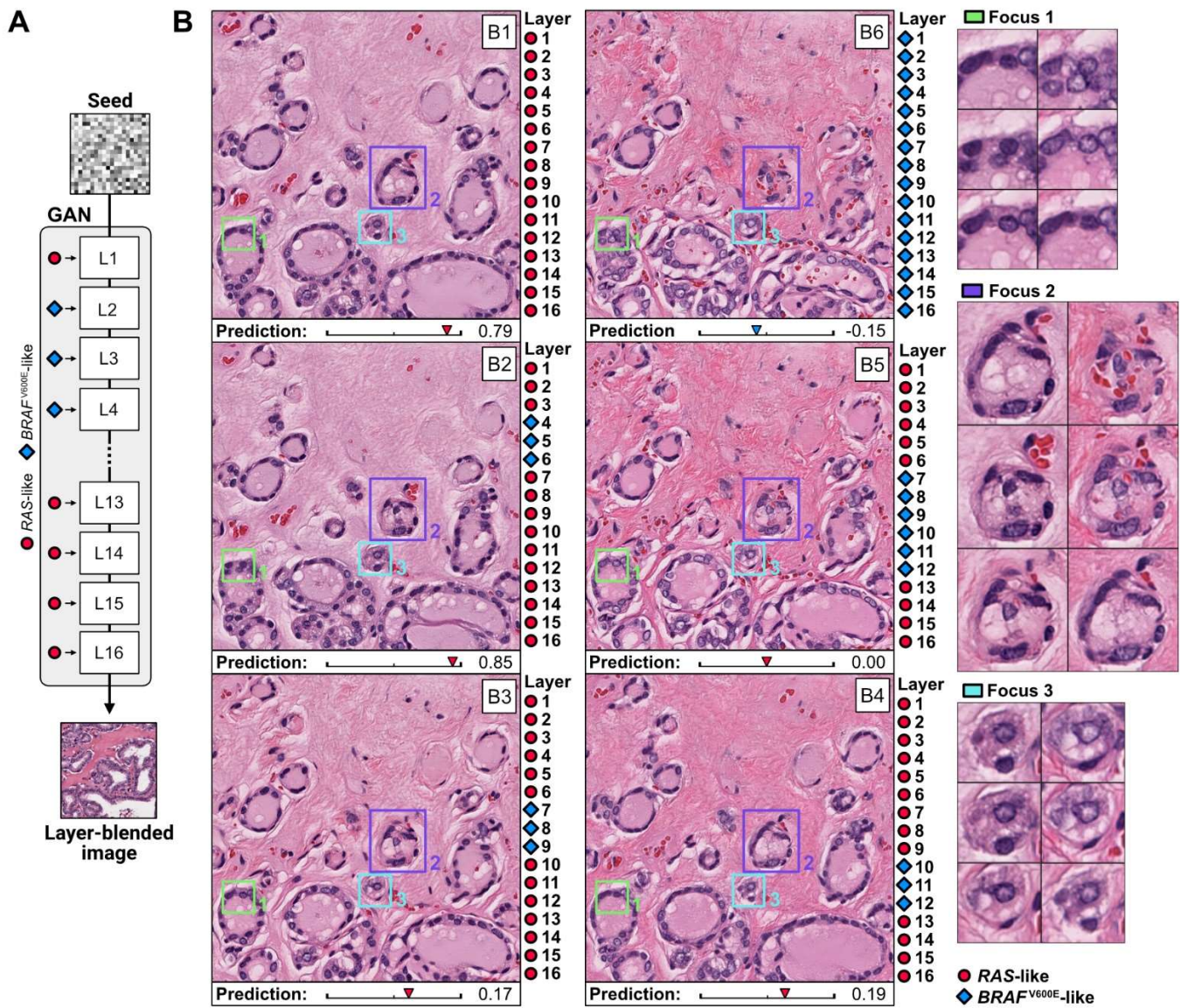

Figure 3

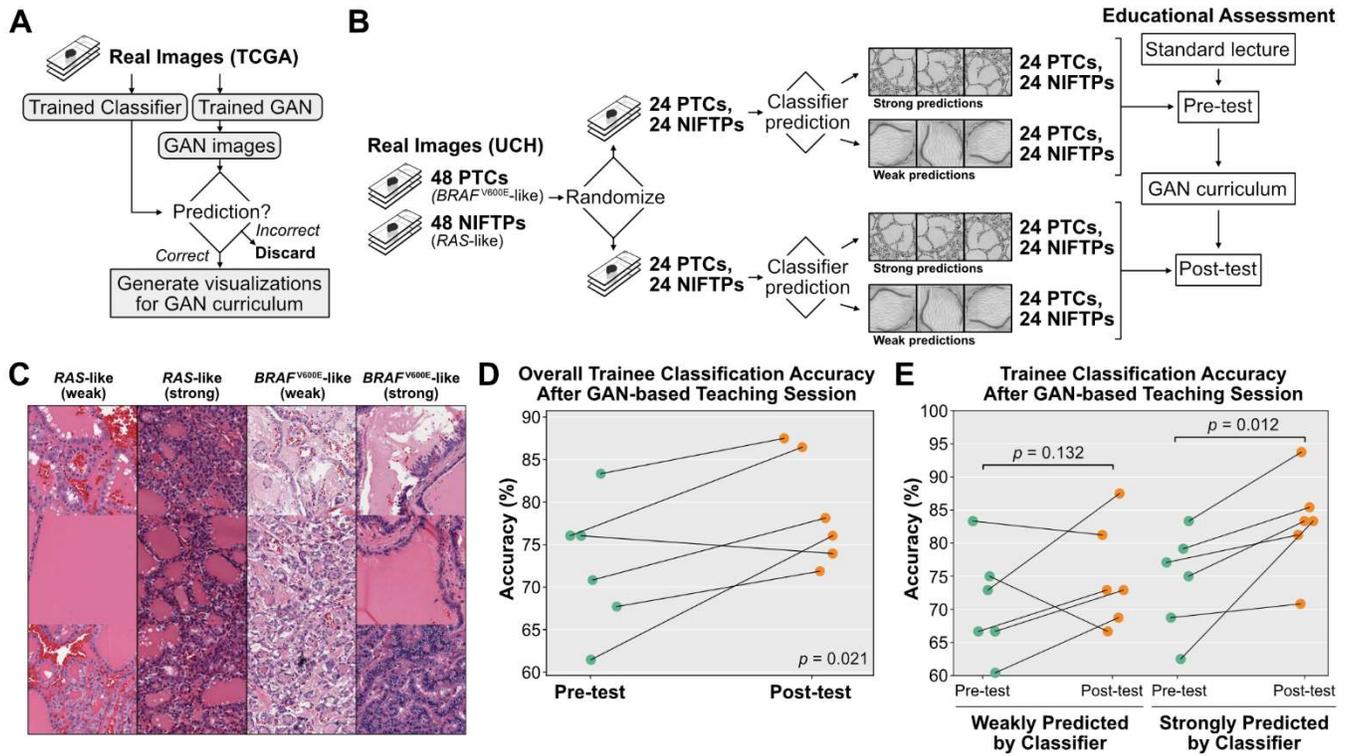

Figure 4